\documentstyle[twocolumn,aps,graphicx]{revtex}

\begin{document}

\title{Pressure Induced Quantum Critical Point and Non-Fermi-Liquid  
Behavior in BaVS$_3$} 
\author{L. Forr\'{o}, R. Ga\'{a}l and H. Berger} 
\address{Institut de G\'eniue Atomique, Ecole Politechnique
Federale de Lausanne, CH-1015 Lausanne, Switzerland}
\author{ P. Fazekas and K. Penc}
\address{Research Institute for Solid State Physics and Optics, 
         H-1525 Budapest, P.O.B. 49, Hungary}
\author{I. K\'{e}zsm\'{a}rki and G. Mih\'{a}ly}
\address{Institut de G\'eniue Atomique, Ecole Politechnique
Federale de Lausanne, CH-1015 Lausanne, Switzerland\\
Department of Physics, Technical University of Budapest, H-1111 Budapest, 
Hungary}

\date{\today}  
\maketitle 

\begin{abstract} 
The phase diagram of BaVS$_3$ is studied under pressure using resistivity 
measurements. The temperature of the metal to nonmagnetic Mott insulator
transition decreases under pressure, and vanishes at the quantum critical
point $p_{\rm cr}=20$kbar. We find two kinds of anomalous conducting
states. The high-pressure metallic phase is a
non-Fermi liquid described by $\Delta\rho\propto T^{n}$ where $n=$1.2--1.3
at 1K$<T<$60K. At $p<p_{\rm cr}$, the transition is preceded by a wide
precursor region with critically increasing resistivity which we
ascribe to the opening of a soft Coulomb gap. 
\end{abstract} 

Understanding the Mott transition, and clarifying the nature of the phases on  
either side of the transition, is a matter of great importance.  
Though metal--insulator transitions are often accompanied  by an ordering  
transition and/or influenced by disorder, one may speak about a ``pure''  
Mott transition which is a local correlation effect in an ideal lattice 
fermion system, 
and takes place without breaking any global symmetry.  Many aspects  
of this problem can be studied on the multifaceted behavior of BaVS$_3$  
\cite{Graf,hall,first}.
  
The metal--insulator transition of the nearly isotropic 3D compound 
BaVS$_3$ offers a realization 
of the pure Mott transition in nature \cite{first}. Under atmospheric pressure 
BaVS$_3$ has  
three transitions: the hexagonal-to-orthorhombic transition at $T_S=240$K  
which has only a slight effect on the electrical properties; the 
metal--insulator transition  
at $T_{\rm  MI}=69$K, which does not seem to break any of the  
symmetries of the metallic phase; and the ordering transition at  
$T_X=30$K \cite{order}.  
In spite of decades of effort, the character of the phases  
and the driving force of the transitions at $T_{\rm  MI}$ and $T_X$, remain  
mysterious. 
 
Here we report the results of single crystal resistivity measurements under  
hydrostatic pressure in the range of 1bar $\le p<$ 25kbar. These pressures  
encompass the entire insulating phase and part of a high-pressure  
low-$T$ conducting phase. We report the first observation of the 
quantum critical point in  
BaVS$_3$, and we characterize the strange metallic phase lying beyond  
the critical pressure $p_{\rm cr}$. On the metallic side of the phase 
boundary, we identify  
two regimes with anomalous properties: (i) a broad region at $p<p_{\rm cr}$  
in which the resistivity increases strongly with decreasing temperature, and  
(ii) a high-pressure non-Fermi-liquid state.  
 
Single crystals of BaVS$_3$ were grown by Tellurium flux method. 
The crystals,  
obtained from the flux by sublimation, have typical dimensions of $3 \times  
0.5 \times 0.5 $ mm$^{3}$. The resistivity was measured in four probe  
arrangement. The current was kept low enough to avoid the self-heating  
of the sample. For the high-pressure measurements the crystal was inserted  
into a self-clamping cell with kerosene as a pressure medium. The pressure 
was  monitored in-situ by an InSb sensor. During cooling down the cell 
there was  
a slight pressure loss, but its influence on the temperature dependence of  
the resistivity was negligible. Above about 15 kbar the pressure was stable  
within 0.1 kbar in the whole temperature range.   
 
 \begin{figure}[ht] 
\centerline{ \includegraphics[width=7.0truecm]{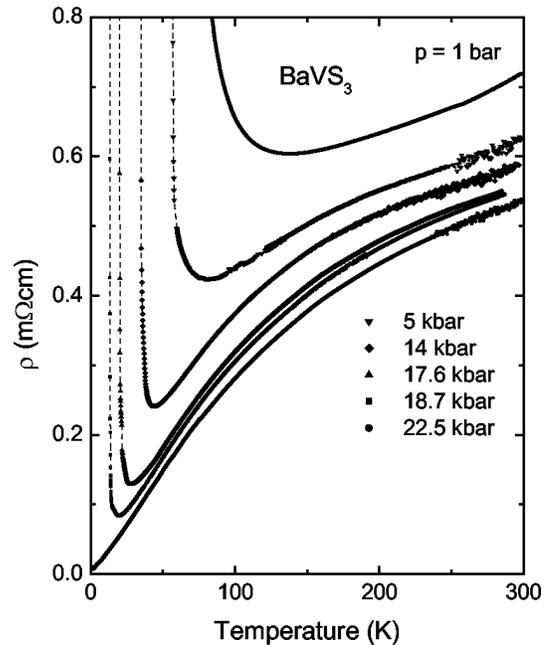} }
\caption{Temperature dependence of the resistivity ${\protect\rho}(T)$  
for various pressures. The $T=0$ insulator-to-metal transition sets in at  
$p_{\rm cr}{\protect \approx} 20{\rm kbar}$.} 
\label{fig:linrt} 
\end{figure}

\begin{figure}[ht] 
\centerline{ \includegraphics[width=7.0truecm]{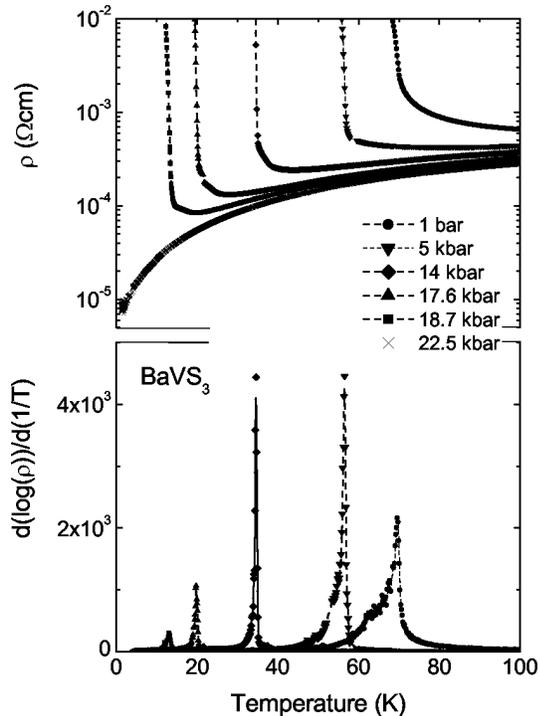} }
\caption{Upper panel: temperature dependence of the resistivity  
$\protect{\log{\rho}}(T)$ for various pressures. Lower part:  
${d\protect\log{\rho}}/d(1/T)$ for selected pressures. The spikes define the  
metal--insulator transition temperatures.} 
\label{fig:logrt} 
\end{figure} 
 
\begin{figure}[ht] 
\centerline{ \includegraphics[width=7.0truecm]{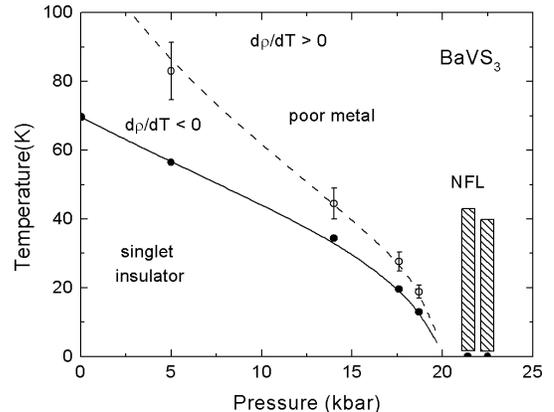} }
\caption{The $p$--$T$ phase diagram of BaVS$_3$ on the basis of resistivity  
measurements. Solid line: metal--insulator phase boundary, 
dashed line: the boundary of the  
precursor region. The columns indicate the range of the non-Fermi-liquid  
$\protect\Delta\rho\sim T^n$ law.} 
\label{fig:phasediag} 
\end{figure}

Figure \ref{fig:linrt} shows the temperature dependence of 
the resistivity for  
various  pressures. As expected from earlier low-pressure data \cite{Graf},  
$T_{\rm  MI}$ decreases smoothly with increasing pressure. The linear plot  
highlights the contrasting behavior of $\rho(T)$ below and above the critical  
pressure, but does not include the regime of higher resistivities. Part of  
this is shown in the logarithmic plot of Fig. \ref{fig:logrt}; it can be  
perceived that the overall resistivity change at the transition remains 
roughly the same, though $T_{\rm MI}$ is suppressed. The pressure dependence  
of the metal--insulator transition temperature was determined from 
the spikes of the logarithmic  
derivative, $d(\log{\rho})/d(1/T)$, as shown for selected pressures in the  
lower panel of Fig. \ref{fig:logrt}. The narrowness of the spikes 
demonstrates  
that the transition remains sharp under pressure. For $p=19.8$kbar we still  
found a metal--insulator transition at $T_{\rm  MI}\approx 5.6$K, but 
for 21.4kbar, the  
resistivity keeps on decreasing at least down to 1K. We estimate  
that $p_{\rm cr}\approx 20$kbar. The phase boundary is shown in Fig.  
\ref{fig:phasediag}. Our resistivity measurements allow the division of the  
conducting phase into further regions of markedly different nature; the  
discussion of these follows.   
 
For $p<p_{\rm cr}$ the resistivity in the metallic phase has a 
marked minimum at  
$T_{\rm min}(p)$ preceding the metal--insulator transition. 
Finding $d\rho/dT<0$ in a metal is  
anomalous, and it is tempting to regard the interval 
$T_{\rm MI}<T<T_{\rm min}(p)$ as an extended precursor regime to 
the insulating phase.   
As shown in Fig. \ref{fig:phasediag} by the dashed line, $T_{\rm min}$ drops  
to zero simultaneously with $T_{\rm MI}$: the insulator and its precursor  
vanish together. We believe that the resistivity minimum is a collective  
effect; if it were due to impurities, it would have no reason to disappear  
beyond $p_{\rm cr}$.  
 
BaVS$_3$ is essentially an isotropic 3-dimensional system  
\cite{first}, thus the appearance of a wide precursor regime within the  
metallic phase is not a regular feature. There is, however, an interesting   
subclass of Mott systems to which it is common: e.g., similar behavior  
is seen above the Verwey transition in magnetite \cite{ti4o7,park}.  
The $T_{\rm min}(p)$ line does not have the significance of a phase boundary;  
it merely marks the temperature where fluctuations 
towards a gapped state become  
so strong that they determine the sign of $d\rho/dT$. The Hall results on  
polycrystalline BaVS$_3$ \cite{hall} imply that the 
number of carriers changes  
in this temperature range, and we believe that resistivity enhancement arises  
from the loss of charge carriers. A phenomenon of this nature is  
observed in Fe$_3$O$_4$, where increasing charge short range order results  
in a diminishing effective number of carriers, and a resistivity minimum  
\cite{park}.  
 
It is remarkable that the apparent opening of a soft charge gap is not 
accompanied  
by the opening of a spin gap; the magnetic susceptibility does not show any  
noticeable anomaly at $T=T_{\rm min}$ \cite{first}. This suggests that the  
phenomenon which sets in at $T_{\rm min}$ is quite distinct from the  
opening of a real gap which happens at $T_{\rm MI}$. $T_{\rm min}(p)$ can  
rather be associated with the onset of charge short range order, and the  
appearance of a soft charge gap which has no effect on the magnetic  
properties. We note that a somewhat similar scenario, but with particular  
emphasis on possible 1D aspects, was considered in \cite{hall}. However,  
the essentially isotropic conductivity \cite{first} rules out 1D  
interpretations.  
   
\begin{figure}[ht] 
\centerline{ \includegraphics[width=7.0truecm]{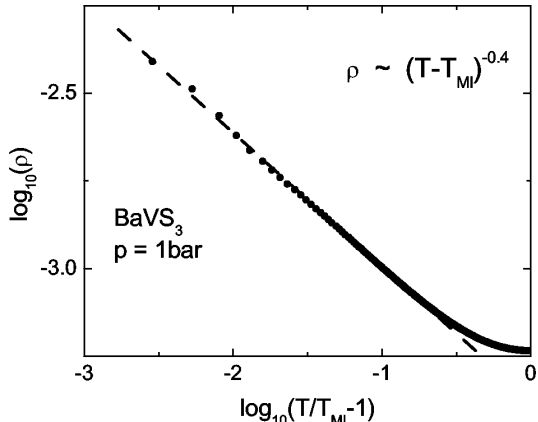} }
\caption{The logarithmic plot of the resistivity vs reduced temperature. The  
resistivity follows a power law in a substantial part of the interval 
$T_{\rm MI}<T<T_{\rm min}$.} 
\label{fig:critic} 
\end{figure}

The resistivity data suggest that the insulating state is  
approached through a regime of critically increasing resistivity. There are  
precedents that valuable insight into the nature of phase transitions in  
strongly correlated systems can be gained by trying to identify critical  
behavior in transport data \cite{kleincr}. The critical behavior can be 
demostrated by  plotting the resistivity, $\rho (t)$, 
as a function of the reduced temperature on logarithmic  
scales ($t=(T-T_{\rm MI})/T_{\rm MI}$). The ambient pressure result is shown  
in Fig. \ref{fig:critic}: the power law $\rho \propto t^{-0.4}$ gives a good  
approximation over more than 30 K above $T_{\rm MI}$, over almost 
2 decades  in the reduced temperature, $t$.

Phenomenologically, low-pressure BaVS$_3$ can be related to other 
(essentially 3D) systems which share at least some of its 
relevant features: the existence of the intermediate disordered insulating 
phase, the resistivity precursor, and the lack of a discernible Fermi edge in 
XPS spectra \cite{naka94}. In addition to magnetite \cite{ti4o7,park} 
we mention Ca$_2$RuO$_4$ \cite{ca2ruo4}, and Ti$_4$O$_7$ 
\cite{ti4o7}. The detailed behavior of either of these systems is quite 
different from that of BaVS$_3$, but we believe that there is also a common
feature: the soft Coulomb gap due to short-range charge fluctuations. 

\begin{figure}[ht] 
\centerline{ \includegraphics[width=7.0truecm]{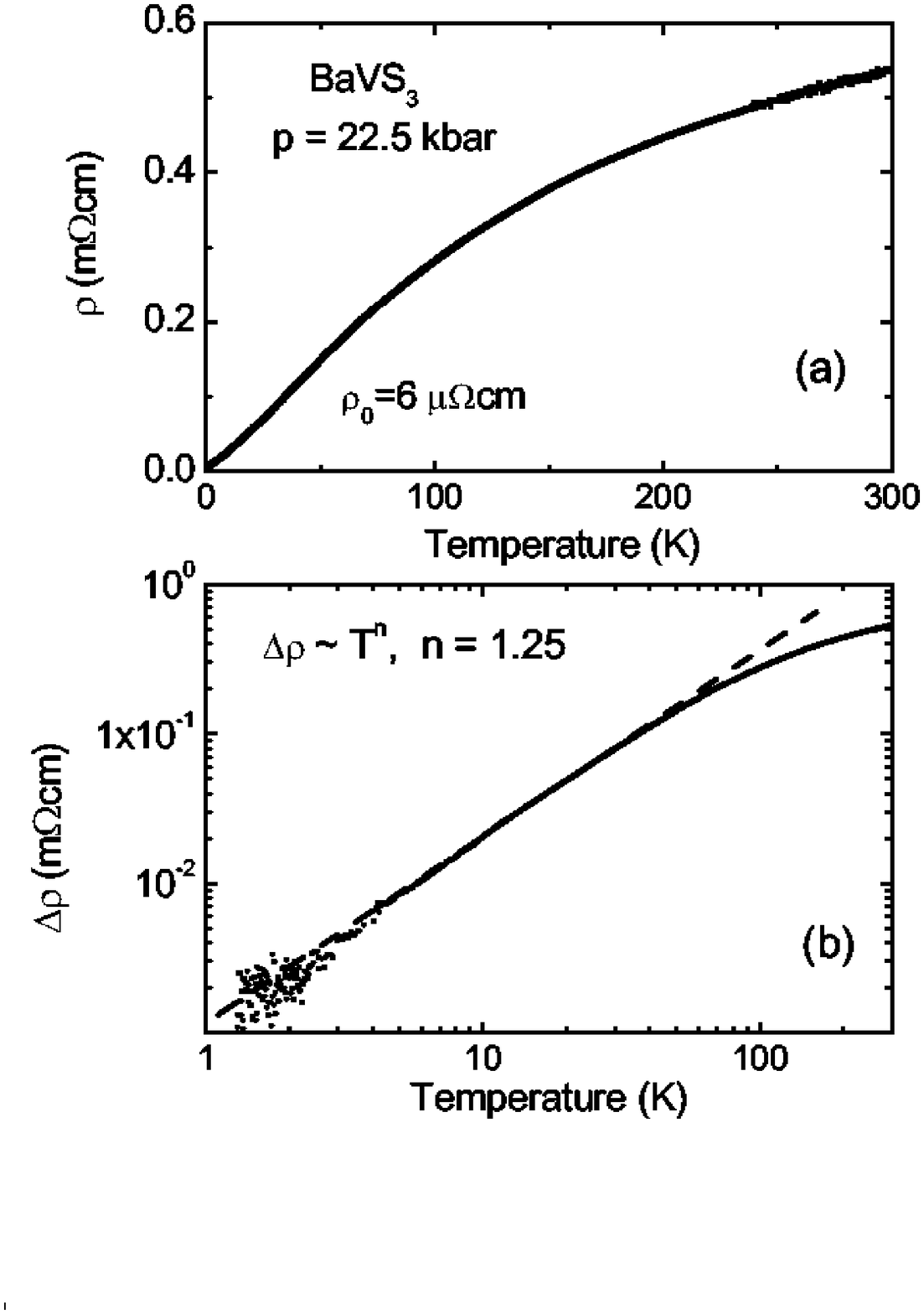} }
\caption{a) The temperature dependence of the resistivity at $p=22.5$kbar 
on linear scales (a), and of $\Delta\rho=\rho-\rho_0$ on logarithmic scales.} 
\label{fig:225rot} 
\end{figure}

Next we discuss the high-pressure metallic phase. Figure \ref{fig:225rot}a  
reveals that the temperature dependence 
of the resistivity is characteristic of  
a bad metal \cite{kivem}. The high temperature behavior is sub-linear,  
and though $\rho(T=300{\rm K})$ corresponds to a mean free path  
$l \sim$5--8{\AA}, which is of the order of the lattice constant, $\rho$  
continues to grow without any sign of saturation. It has been shown that  
strong electron--phonon scattering could account for such a behavior  
\cite{millis}, but we believe that in our case the electron--electron  
scattering dominates. This assumption is supported by the specific heat  
data \cite{Imai}: the electronic (primarily orbital) entropy keeps on  
increasing even beyond 300K. The magnitude and the unusual shape of  
$\rho (t)$ indicate a new scattering process, which is to be associated  
to orbital fluctuations. 
  
The low temperature region of the pressure-induced metallic phase is  
particularly interesting. For $T<60$K the resistivity does not follow  
the characteristic Fermi liquid 
behavior $\Delta\rho=\rho(T)-\rho_0\propto T^2$  
($\rho_0$ is the residual resistivity).  
Furthermore, the log--log plot of $\Delta\rho(T)$ is approximately linear  
in an extended range, allowing to fit the resistivity with the customary  
non-Fermi-liquid `law' $\Delta\rho\propto T^n$, where in general $1\le n<2$  
\cite{sbarbar}. Figure \ref{fig:225rot}b shows that $\Delta\rho\propto  
T^{1.25}$ gives an excellent description for BaVS$_3$ for 1K$<T<$50K.  
A somewhat larger temperature range 
with $n\approx 1.2$ is found at $p=21.4$kbar. 
The extent of the non-Fermi-liquid regime is given by columns in Fig.  
\ref{fig:phasediag}.

The above low temperature behavior is similar 
to that of nearly antiferromagnetic  
$f$-electron systems such as CePd$_2$Si$_2$ \cite{mathur}.  
The interpretation usually invokes nearness to a quantum 
critical point, or the existence  
of rare regions \cite{sbarbar,qcp,voj99a}. In these cases the 
non-Fermi-liquid region  
is placed into a phase diagram where the static properties of the phases  
are in principle well understood. It is not so with BaVS$_3$ for which there  
is no consensus either about the driving force of the metal--insulator 
transition, or the  
nature of the low-$T$ phases. Since even weak extrinsic disorder can have  
a drastic effect on the critical behavior, the relative importance of  
correlation and disorder should also be considered both for the insulator  
and the various conducting regimes. 
  
The effects of the vicinity of a ferromagnetic, or an antiferromagnetic, 
quantum critical point on the  
metallic resistivity have been worked out \cite{sbarbar}. The overall 
appearance of the  
susceptibility curve \cite{first} shows that BaVS$_3$ is 
dominated by antiferromagnetic  
spin--spin interactions, so the predictions concerning the resistivity of  
a nearly antiferromagnetic metal are relevant. It has been argued 
that for samples  
of sufficiently good quality, $\Delta\rho\propto T^n$ where $n<1.5$, and  
finding $n=1.2$--1.3 over 1--2 decades of $T$ is a reasonable expectation  
\cite{rosch}. This is in full accordance with our results. The fact that at  
$p=22.5$kbar $\rho_{300}/\rho_0\sim 100$, shows that our sample is of good  
quality and disorder effects are weak as far as the high-pressure conductor  
is concerned. 

Let us note here that CaRuO$_3$ provides another example of a $d$-electron 
system whose non-Fermi-liquid nature is probably explained by its being nearly 
antiferromagnetic. However, its resistivity follows 
the $\Delta\rho\propto T^{1.5}$ 
relationship \cite{caruo3}, which is expected for dirty samples \cite{rosch}. 
 
The reason for the lack of magnetic long range order in the 
$T_X<T<T_{\rm MI}$ insulating phase   
is not evident. One may first think that the in-plane frustration is  
responsible because the V ions form triangular $a$--$b$ planes. However,  
neither the isotropic nor the anisotropic triangular Heisenberg model is,  
for any spin, frustrated enough to give a spin liquid \cite{kleine}.  
Here one may be tempted to invoke disorder: it is known that   
quantum antiferromagnets (especially for $S=1/2$) tend to be unstable 
against the formation of a  
random singlet phase \cite{bl82}. The theoretical issue is still open,  
but the outcome that antiferromagnetism should be always unstable against  
quenched disorder, is considered implausible \cite{voj99a}.  
 
We believe that in a weakly frustrated system like BaVS$_3$, the Heisenberg  
model would order, and BaVS$_3$ is non-magnetic because  
its effective hamiltonian is not a pure Heisenberg model. We argued in  
Ref. \cite{first} that including the orbital degrees of freedom of the  
low-lying crystal field quasi-doublet, one finds a large number of  
energetically favorable dimer coverings of the triangular lattice such  
that intra-dimer spin coupling is strong, while inter-dimer perturbations are  
weak. The equilibrium phase at $T_X<T<T_{\rm MI}$ can be visualized as a  
thermal average over a class of valence bond solid states. Intra-dimer  
interaction causes the opening of a spin gap, while inter-dimer interactions  
results intermediate-range correlation, and a {\bf Q}-dependence in the spin  
excitation spectrum \cite{spingap}. In this model we do not have to invoke  
disorder to explain the singlet phase. Quite on the contrary: either sulfur  
off-stoichiometry \cite{shiga98}, or Ti-doping \cite{matsu}, is known to  
break up singlet pairing. The well-developed singlet insulating phase is the  
hallmark of a clean system.  
   
In conclusion, we studied the phase diagram of the non-magnetic Mott system  
BaVS$_3$ by resistivity measurements.  A quantum critical 
point $p_{\rm cr}\approx  
20$kbar was found, and the results revealed the existence of anomalous  
conducting regimes. The pressure-induced metallic phase is a 
non-Fermi-liquid at  
$T<60$K. We suspect that at high temperatures the electrical transport  
is determined by the scattering on orbital fluctuations. For $p<p_{\rm cr}$  
an unusually wide precursor regime was identified above the 
metal--insulator transition.  
In this regime, which is attributed to the appearance of a soft charge gap,  
$\rho (t)$ can be well described by a power law of the reduced temperature.  
The microscopic nature of these regimes remains to be elucidated.    
 
This work was supported by the Swiss National Foundation for Scientific  
Research and by Hungarian Research Funds OTKA T025505 and D32689,  
FKFP 0355 and B10, AKP 98-66 and Bolyai 118/99.

\end{document}